# A simple predictor based on delay-induced negative group delay

Henning U. Voss

Weill Cornell Medical College, Citigroup Biomedical Imaging Center
516 East 72nd Street, New York, NY10021, USA
e-mail address: hev2006@med.cornell.edu

A very simple linear signal predictor that uses past predicted values rather than past signal values for prediction is presented. Man-made or natural systems utilizing this predictor would not require a memory of input signal values but only of already predicted, internalized states. This delay-induced negative group delay (DINGD) predictor affords real-time prediction of signals without the need for a specific signal model. Its properties are derived analytically and are numerically tested on various types of broadband input data.

## Introduction

Negative group delay (NGD) of an input/output system causes the output signal to anticipate or predict characteristics of the input signal. NGD and the related concept of negative group velocity have been theorized and experimentally found in systems with anomalous dispersion [1-4], metamaterials [5-7], transmission lines [8, 9], and electronic circuits [10-13]. Recently, it has been shown that negative group delay can also occur in continuous-time systems with time-delayed feedback, or mathematically, non-autonomous delay-differential equations [14]. Time delays are a typical component of biological neuronal networks, and it is reasonable to hypothesize a possible relevance of this *delay-induced negative group delay* (DINGD) mechanism in neuronal computations [15] involved, for example, in human motor control [16].

The pioneering paper of Mitchell and Chiao [17] experimentally demonstrated NGD for Gaussian waveforms in an electronic circuit and also showed that causality is retained. They used the concept of transfer functions, from which the frequency dependent group delay can be derived for any input signal waveform independent of its shape. Therefore, some systems with NGD can be viewed as real-time signal predictors [18], which can be understood by analyzing their transfer or frequency response functions. The absolute value of the group delay then defines the prediction horizon, i.e., the time the output y(t), here called a predictor, predicts the input x(t) ahead of time.

In this tutorial-style manuscript very simple, probably the simplest possible, DINGD predictors are introduced. They are given by discrete-time systems, which simplifies numerical simulations and would allow for digital signal processing implementation. They are still delay-induced NGD predictors in the following sense: It will turn out that these predictors do not use past input signal values x(t-1), x(t-2), ... for prediction, as most conventional predictors do, but only previously predicted output values y(t-1), y(t-2), ..., along with the present input value x(t). The previously predicted output values are delayed feedback inputs to the predictor. This scheme could have advantages in natural or man-made applications.

In the following, discrete-time DINGD predictors will be described, theoretically analyzed, and their performance will be illustrated with various numerical simulations of real-time, broadband signal prediction.

## The DINGD predictor

The simplest DINGD predictor is defined as the discrete-time non-autonomous linear system

$$y(t) = bx(t) - cy(t - \tau), \tag{1}$$

where x(t) is a scalar input signal whose forthcoming values are to be predicted by y(t), b a non-zero input scaling parameter, c a non-zero delayed feedback gain, and $\tau$ a positive integer, a time delay. Time is restricted to multiples of a sampling time interval $\Delta t$, i.e., t = ..., -$\Delta t$, 0, $\Delta t$, 2$\Delta t$, .... For simplicity, we set $\Delta t$ = 1 in the following, such that t = ..., -1, 0, 1, 2, ... and $\tau$ = 1, 2, ....

In order to understand how the discrete-time DINGD predictor predicts, it is necessary to derive its frequency-dependent group delay. Although Eq. (1) looks quite simple, it was not possible for me to derive its prediction properties by any other, more intuitive, means. (An attempt has been made for the time-continous analogue of Eq. (1) in Ref. [14], where a heuristic explanation was provided to relate its behavior to anticipatory synchronization [19], called „anticipatory relaxation dynamics". In Ref. [20] it had been conjectured that anticipatory synchronization is related to the findings of Mitchell and Chiao but it was not specified how exactly.)

The frequency response function defines the input/output relationship between x(t) and y(t) under steady-state conditions in Fourier space as

$$Y(\omega) = H(\omega)X(\omega),$$

where $\omega = 2\pi f$ is the frequency in rad/time, f is frequency in oscillations/time, $x(t) = \int X(\omega)e^{i\omega t}d\omega$, and $y(t) = \int Y(\omega)e^{i\omega t}d\omega$. The latter two expressions are inverse Fourier transforms; the sign convention here is opposite to Ref. [17], following the majority of the literature.

The frequency response function can be written in terms of phase and gain as

$$H(\omega) = |H(\omega)|e^{i\Phi(\omega)}.$$

The frequency response function of the discrete-time DINGD predictor (1) can be found by inserting the inverse Fourier components of x and y into the predictor. It is

$$H(\omega) = \frac{b}{1 + ce^{-i\omega\tau}} = \frac{b}{\beta(\omega)}(1 + c\cos(\omega\tau) + i\,c\sin(\omega\tau)) \tag{2}$$

with

$$\beta(\omega) = 1 + 2c\cos(\omega\tau) + c^2.$$

Its gain is

$$|H(\omega)| = \frac{|b|}{\sqrt{\beta(\omega)}},$$

and its group delay

$$\delta(\omega) = -\frac{d\Phi(\omega)}{d\omega} = -\frac{c\tau(c + \cos(\omega\tau))}{\beta(\omega)}. \tag{3}$$

The latter expression is properly defined outside of the poles of the frequency response function only.

In order to make predictions one sample step or more ahead, we are seeking to obtain an integer group delay $\delta \leq -1$ for $\tau \geq 1$ (causal system) and $0 < c \leq 1$ (to avoid instability). It is most instructive to first consider the zero frequency case ($\omega = 0$) and from there to derive the group delay for general frequencies. The group delay for zero frequencies is

$$\delta(0) = -\frac{c\tau}{1 + c}. \tag{4}$$

Specific cases for the time delay $\tau$ are considered:

• $\tau = 1$: There is no integer solution for $\delta(0)$ for any $0 < c \leq 1$. (This is a special case of the elementary NGD IIR filter introduced by Ravelo [21] ($b = 0$ in Eq. (1) there, $T_s = 1$).)

• $\tau = 2$: For $c = 1$ the group delay at zero frequency is $\delta(0) = -1$. Furthermore, the group delay is -1 for all frequencies where it is defined, i.e., outside of the poles of the frequency response function. The frequency response function has poles defined by the zeros of its denominator. The poles are located at $\omega = (2n+1)\pi/2$ ($n = 0, 1, ...$). The highest frequency for a signal sampled with $\Delta t = 1$ is $\omega_N = \pi$, so there is one pole, at $f_1 = \omega_1/2\pi = ¼$. This pole separates two frequency bands with qualitatively different properties: Due to a phase jump at $f_1$ the second band causes a sign reversal of the output and thus cannot be used together with the first band for prediction.

Figure 1 shows in the upper left panel the frequency response function expressed through its gain and phase, as well as the group delay over frequency for this case. From this figure it is clear that the DINGD predictor depends, as all NGD based prediction, on the frequency content of the signal to be predicted.

• $\tau = 4$: For $c = 1$ the group delay at zero frequency is $\delta(0) = -2$. Furthermore, the group delay is -2 for all frequencies for which it is defined. The poles of the frequency response function are located at $\omega = (2n+1)\pi/4$ ($n = 0, 1, ...$). There are two poles, at $f_1 = \omega_1/2\pi = 1/8$ and $f_2 = \omega_2/2\pi = 3/8$. These two poles separate three frequency bands with qualitatively different properties: The first band and the third band can be used combined. The second band causes prediction with reversed sign and cannot be combined with the other two bands. However, this band could also be useful, too. To use it, one can set $b < 0$ in Eq. (1) to compensate for the sign reversal.

Figure 1 shows in the lower left panel the frequency response function and group delay.

- Larger, even τ: It is straightforward to generalize to larger delays as long as c = 1. In general, as the delay increases, there will be more poles. This means the frequency bands will be split up into more sections. Also, in general the prediction horizon will always be half of the predictor feedback delay τ as per Eq. (4).

In order not to cause instability of system (1) and also in physical implementations a value of c = 1 might not be feasible and it is more useful to proceed with c = 1 - ε, with ε ≪ 1. This has the additional advantage that the poles of the frequency response function are resolving into mere resonances (i.e., the gain is not diverging). Figure 1 shows in the right two panels the frequency response function as well as the group delay for c = 0.99 and τ = 2, 4.

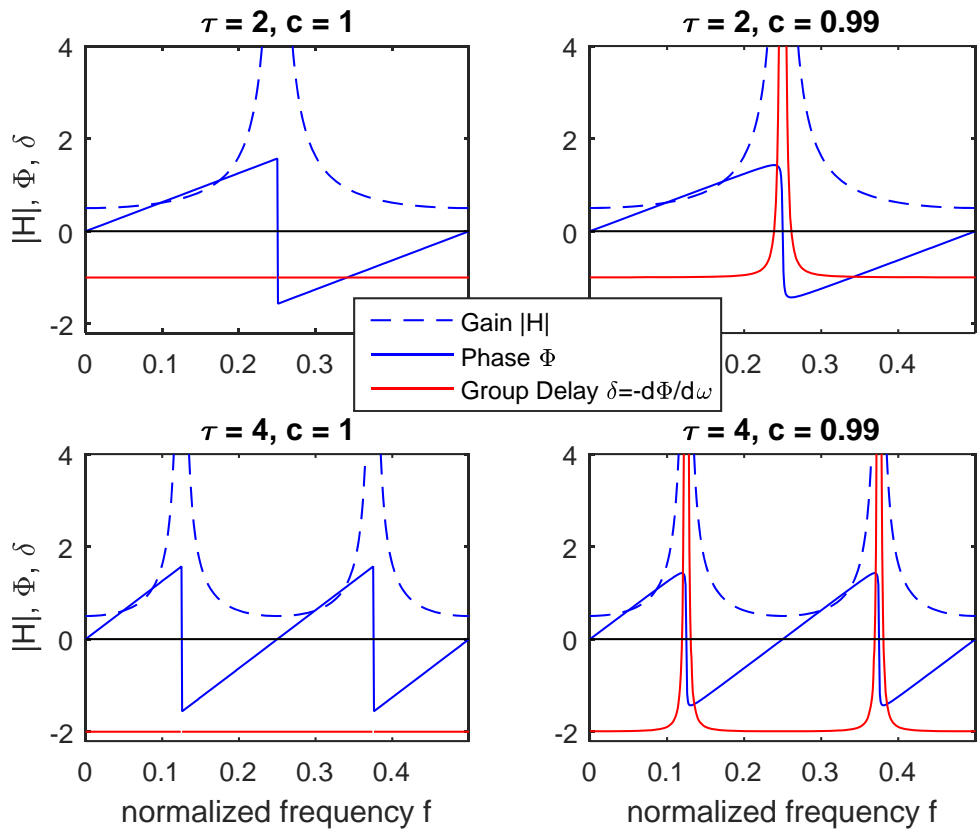

Figure 1: **Theory I.**
Gain and phase of frequency response function (2) as well as corresponding group delay (3) for feedback delays τ = 2 (top row) and τ = 4 (bottom row), for feedback gain c = 1 (left column) and c = 0.99 (right column). The meaning of the graphs is shown in the legend. Note that although the NGD is constant over a wide frequency range, the gain increases for high frequencies, thus restricting applicability of the DINGD predictor to signals containing only lower frequencies. This is demonstrated in the applications in Figs. 2 to 5.

## Application examples

The performance of the discrete-time DINGD predictor will be illustrated with the help of four examples, Figs. 2 to 5. The examples are described in the figure captions. In all examples, c = 1 - ε (ε ≪ 1) is used. All computations were performed with MATLAB R2015a (The MathWorks, Inc., Natick, MA).

## Figure 2: **Band-limited noise I.**

First row: A band-limited noise signal x(t) (black, thick line) and its prediction signal y(t) (red), the output of the predictor (1). Out of 1000 simulated time points, 100 are shown.

Second row, left image: The cross-correlation function CCF(τ) between x(t) and y(t) peaks at a lag of -1. Its peak value is CCF(-1) = 0.98. This shows that y(t-1) ≈ x(t), or, equivalently, y(t) ≈ x(t+1). Therefore, y(t) is a predictor of x(t). Center and right image: Scatterplots between x(t) and y(t) and between x(t) and y(t-1). These plots confirm that y(t) is more correlated with x(t+1) than with x(t), although the DINGD predictor, Eq. (1), depends on x(t) only and not on x(t+1).

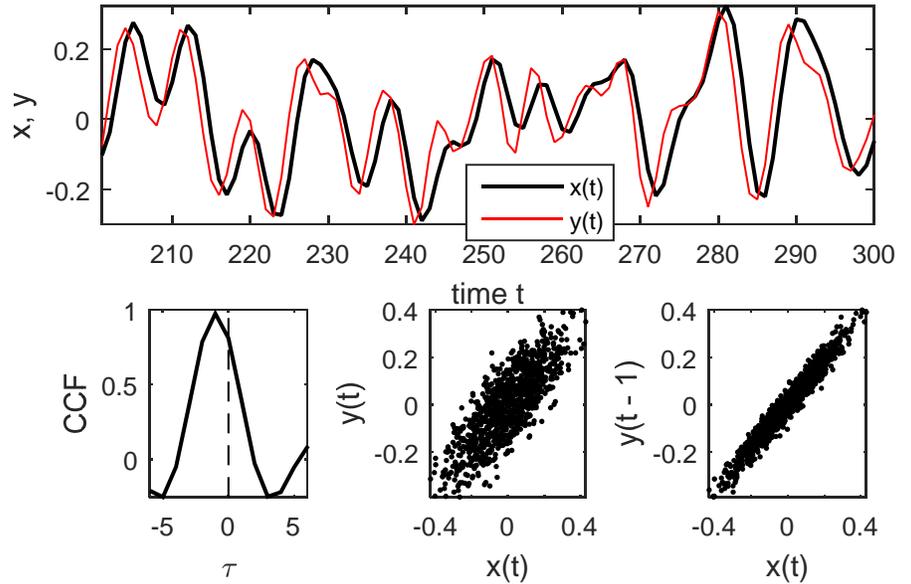

Third row: Power spectral density function estimates for x(t) (black, thick line) and y(t) (red). One can see that frequency components of Y(ω) that are closer to the resonance, where the gain increases (see fourth row), are amplified more, which causes the more jittery appearance of the predictor time series y(t) as compared with the signal time series x(t) in the first row.

Fourth row: Gain and phase of the frequency response function as well as corresponding group delay. Shown are analytic values from Eqs. (2) and (3), as well as gain and phase values estimated from the data, as shown in the legend.

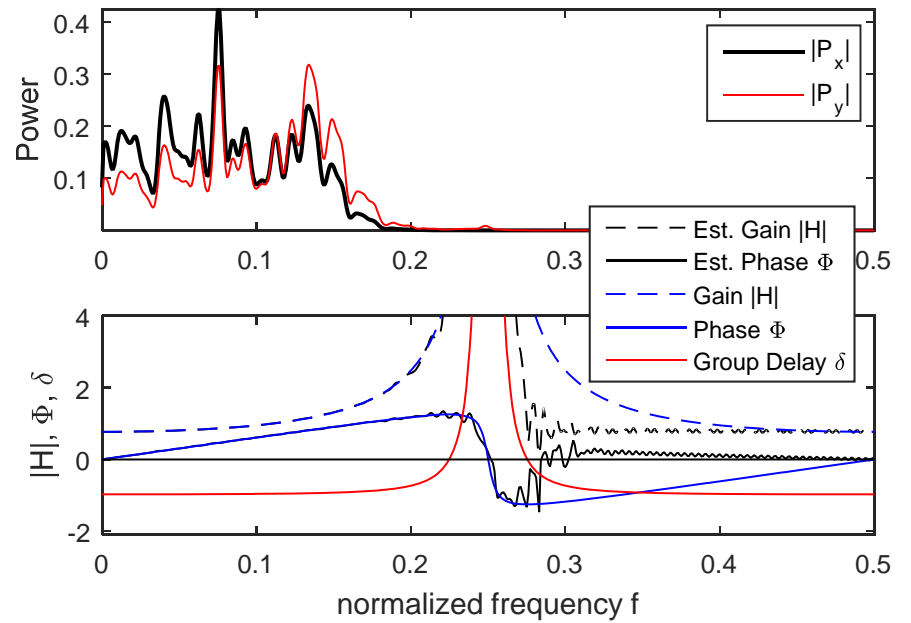

Signal and prediction parameters: The signal x(t) consists of 1000 samples of white noise, low-pass filtered with a Butterworth filter of seventh order with a cutoff frequency of 0.15. The parameters of Eq. (1) are b = 1.50, c = 0.95, and τ = 2.

Figure 3: **Chirp signal I.**

First row: The left plot shows the first 200 data points of a chirp signal (a frequency-swept sine function) x(t) and its prediction signal y(t). The right plot shows the last 20 data points. Whereas an enlargement of the left plot would show prediction (with reduced amplitude), too, prediction is more evident in the right plot with the higher frequency oscillations. The DINGD predictor predicts the signal with the same group delay of -1, for both very low and high frequencies. This is an example for a non-stationary signal that still can be predicted in real time.

Second row, left: The cross-correlation function has a peak value of CCF(-1) = 0.99. This shows that y(t-1) ≈ x(t), or, equivalently, y(t) ≈ x(t+1). Therefore, y(t) is a predictor of x(t). Center and right: The scatterplots confirm that y(t) is more correlated with x(t+1) than with x(t).

Third row: Power spectral density function estimates for x(t) (black, thick line) and y(t) (red). It is evident that frequency components of the signal that are closer to the resonance of the frequency response function (fourth row), where its gain increases, are amplified more relative to lower frequency components. This explains the relatively smaller amplitude of the predictor for low frequencies. In other words, the parameter b has been set to provide correct amplitudes for high frequencies only. The large estimation errors of gain and phase for frequencies > 0.15 are due to the lack of input signal power for those frequencies.

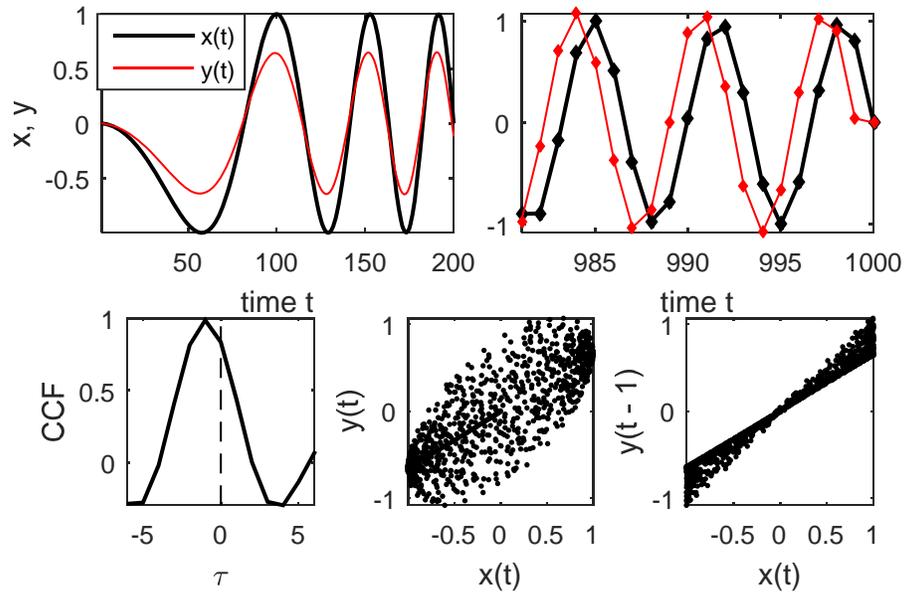
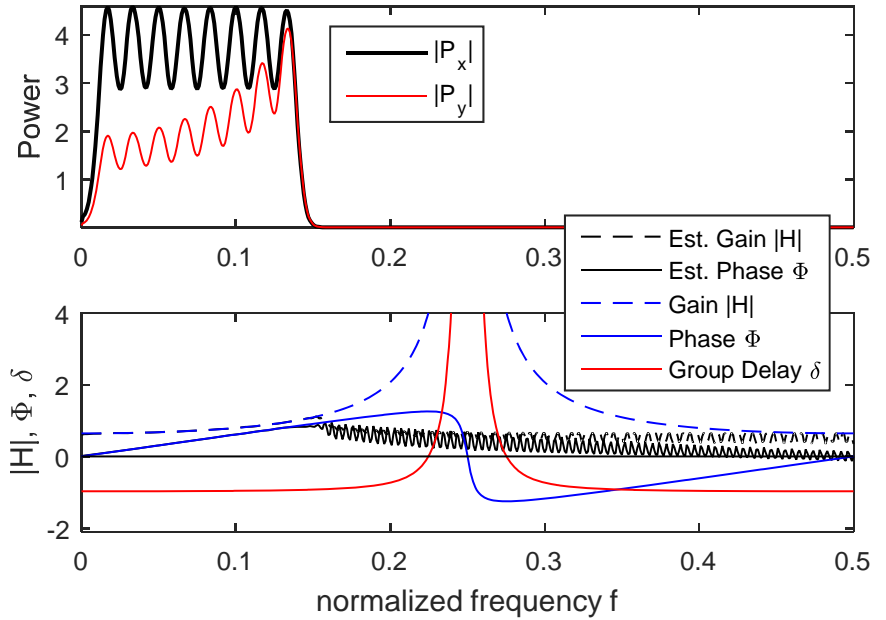

Fourth row: Gain and phase of the frequency response function as well as corresponding group delay. Shown are analytic values from Eqs. (2) and (3), as well as gain and phase values estimated from the data, as shown in the legend.

Signal and prediction parameters: The signal x(t) consists of 1000 samples of a linear chirp signal with cutoff frequency of 0.15. The parameters of Eq. (1) are b = 1.25, c = 0.95, and τ = 2.

**Figure 4: Band-limited noise II – Prediction with the second NGD band and a group delay of -2.**

First row: A band-limited noise signal x(t) (black, thick line) and its prediction signal y(t) (red), the output of the predictor (1). Out of 1000 simulated time points, 50 are shown. It is evident that the signal is predicted two time steps ahead. It is also evident that, although it is not smooth, the signal is predicted with high accuracy, including patterns of data points that are not formed by the envelope of an oscillatory signal.

Second row, left image: The cross-correlation function CCF(τ) between x(t) and y(t) peaks at a lag of -2. Its peak value is CCF(-2) = 0.98. This shows that y(t-2) ≈ x(t), or, equivalently, y(t) ≈ x(t+2). Therefore, y(t) is a predictor of x(t). Center and right image: Scatterplots between x(t) and y(t) and between x(t) and y(t-2). These plots confirm that y(t) is more correlated with x(t+2) than with x(t).

Third row: Power spectral density function estimates for x(t) (black, thick line) and y(t) (red).

Fourth row: Gain and phase of the frequency response function as well as corresponding group delay. Shown are analytic values from Eqs. (2) and (3), as well as gain and phase values estimated from the data, as shown in the legend. Again, outside of the signal frequency band the estimates naturally have a large error.

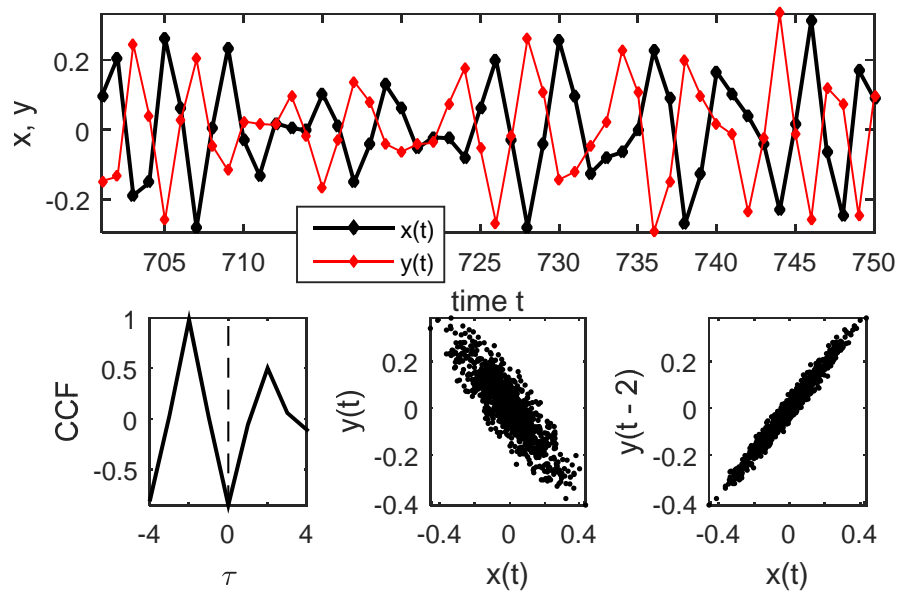
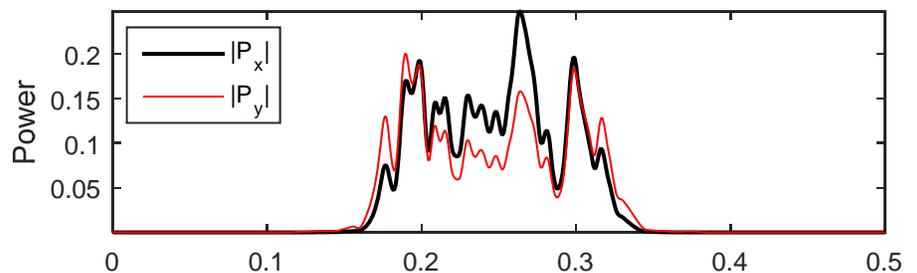
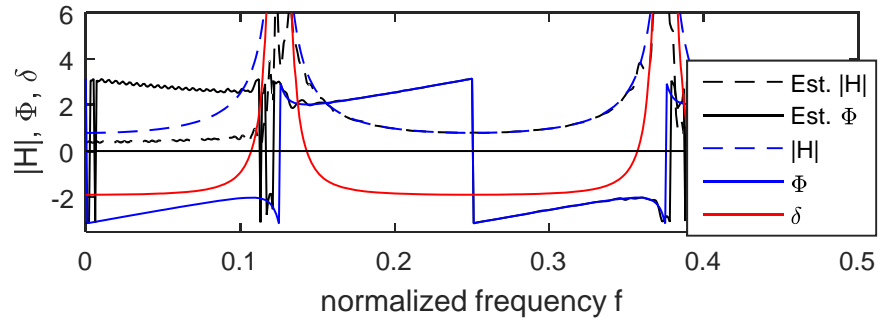

Signal and prediction parameters: The signal x(t) consists of 1000 samples of white noise, band-pass filtered with a Butterworth filter of seventh order with cutoff frequencies of 0.18 and 0.32. The parameters of Eq. (1) are b = -1.50, c = 0.90, and τ = 4. Note that the phase behavior of the second NGD band relative to the first band requires a negative value for the parameter b.

Figure 5: **Neuronal signal, predicted with a group delay of -8.**

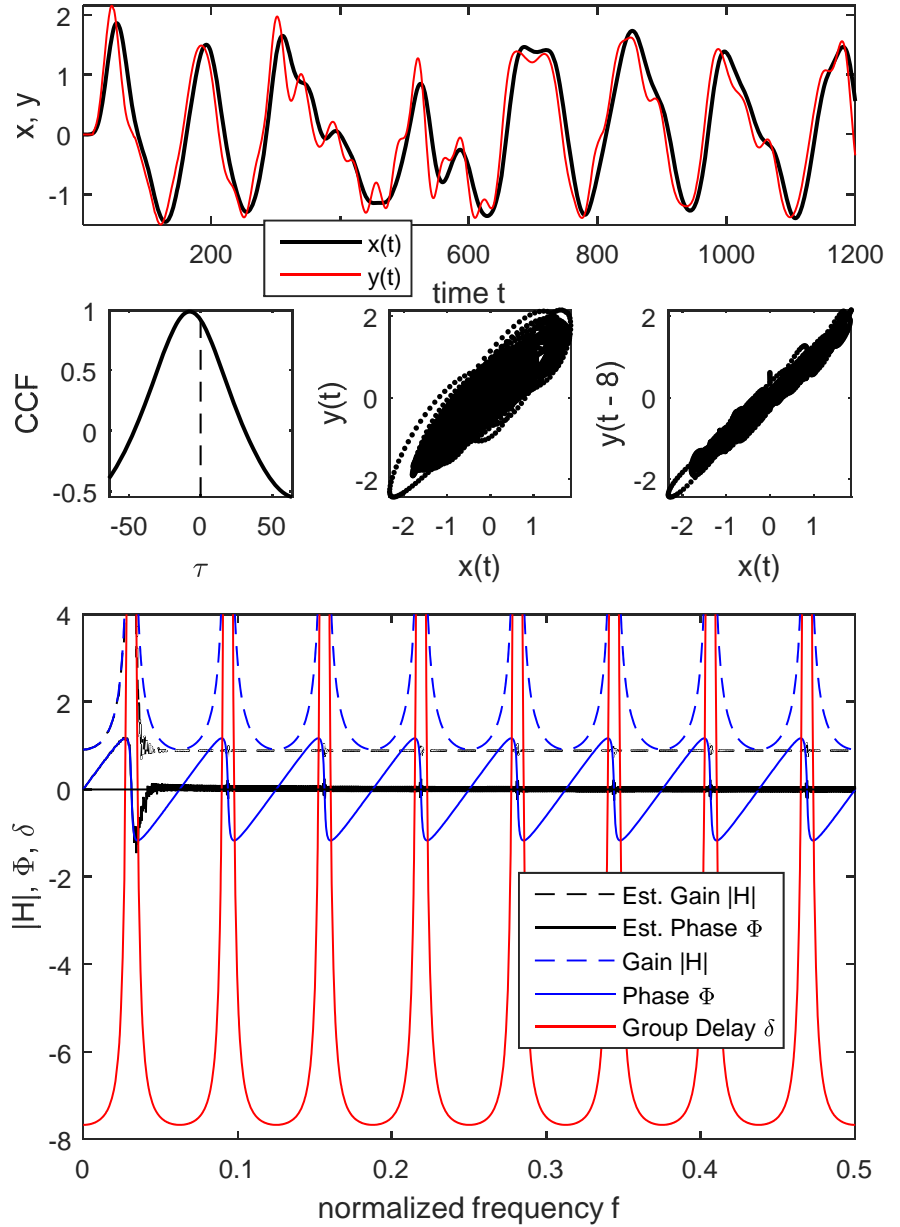

First row: The neuronal signal $x(t)$ (black, thick line), a local field potential from the left hippocampus of a rat, obtained from CRCNS.org [22] and filtered as described in Ref. [15]. Here 1200 out of 6250 used data points are shown. The prediction signal $y(t)$ (red) predicts the input $x(t)$ eight time steps ahead, with an occasional small oscillatory error.

Second row, left image: The cross-correlation function $CCF(\tau)$ between $x(t)$ and $y(t)$ peaks at a lag of -8. Its peak value is $CCF(-8) = 0.99$. A group delay of $\delta = -8$ corresponds to a time of 6.4 ms in the original time scale of the data. In Ref. [15] this signal had been predicted with a group delay of -7.2 ms by using a more specific model based on delayed-leak integrators, which also can have NGD, caused by the mechanism of anticipatory relaxation dynamics [14]. Center and right image: Scatterplots between $x(t)$ and $y(t)$ and between $x(t)$ and $y(t-8)$.

Third row: Gain and phase of the frequency response function as well as corresponding group delay. There are now $\tau/2 + 1 = 9$ distinct frequency bands with NGD. Due to the choice of $c = 0.92$ the value of $\delta = -8.00$ is not completely attained; however, in practice the signal is predicted with a prediction horizon of 8 due to the integer sampling time.

Prediction parameters: The parameters of Eq. (1) are $b = 1.75$, $c = 0.92$, and $\tau = 16$.

# Cascading

As it has been shown above, one way to increase the NGD is by increasing the feedback delay $\tau$ of the DINGD predictor. But there is another way: Feeding the output of the predictor into another predictor, or "cascading" [12, 13, 23] predictors. This way, it is possible to increase the NGD without increasing the feedback delay. This has the advantage that one can work with a delay of $\tau = 1$; although the NGD without cascading would be 0.5, as we have seen before, with cascading one can again obtain integer values.

Cascading $m > 1$ DINGD predictors in the time domain means that the output $y(t) = y_m(t)$ relates to the input $x(t)$ via

$$y_1(t) = bx(t) - cy_1(t-\tau),$$
$$y_2(t) = by_1(t) - cy_2(t-\tau),$$
$$\ldots$$
$$y_m(t) = by_{m-1}(t) - cy_m(t-\tau).$$
(5)

The frequency response function is

$$H_m(\omega) = H(\omega)^m = |H(\omega)|^m e^{im\Phi(\omega)}.$$
(6)

Its gain is

$$|H_m(\omega)| = \frac{|b|^m}{\beta(\omega)^{m/2}},$$

and its group delay

$$\delta_m(\omega) = -\frac{d\Phi_m(\omega)}{d\omega} = m\delta(\omega),$$
(7)

defined wherever $\delta(\omega)$ is defined.

We consider the special case of $m = 2$, $\tau = 1$ first. Again, this is a special case of the IIR filter considered in Ref. [21], this time with cascading. As before, in order to make predictions one sample step or more ahead, we need to obtain an integer group delay $\delta \leq -1$. Since the group delay for zero frequencies is just m times the group delay for the case of no cascading, for $c = 1$ the group delay at zero frequency is $\delta_2(0) = -1$. Furthermore, for $c = 1$ the group delay is constant -1 for all frequencies where it is defined. The poles of the frequency response function are located at $\omega = (2n+1)\pi$ ($n = 0, 1, \ldots$). Therefore, there is only one pole at $f_1 = \omega_1/2\pi = ½$, at the edge of the frequency range.

In summary, for $\tau = 1$ and $m = 2$ the group delay is negative throughout the entire frequency range. This means that cascading allows for using a delay of $\tau = 1$ without a split of frequency bands, and the NGD can, in the ideal case, extend over the entire frequency band. Still, the variable gain of the frequency response function prevents the prediction of signals with certain frequencies.

It is straightforward to derive cases with $\tau > 1$ and $m > 2$ etc. However, the case of $\tau = 1$ stands out as the NGD bands are not split up into separate bands.

Figure 6 shows examples for m = 2 and m = 4, again for c = 1 and c = 1 - ε, with ε ≪ 1. Figure 7 provides an example.

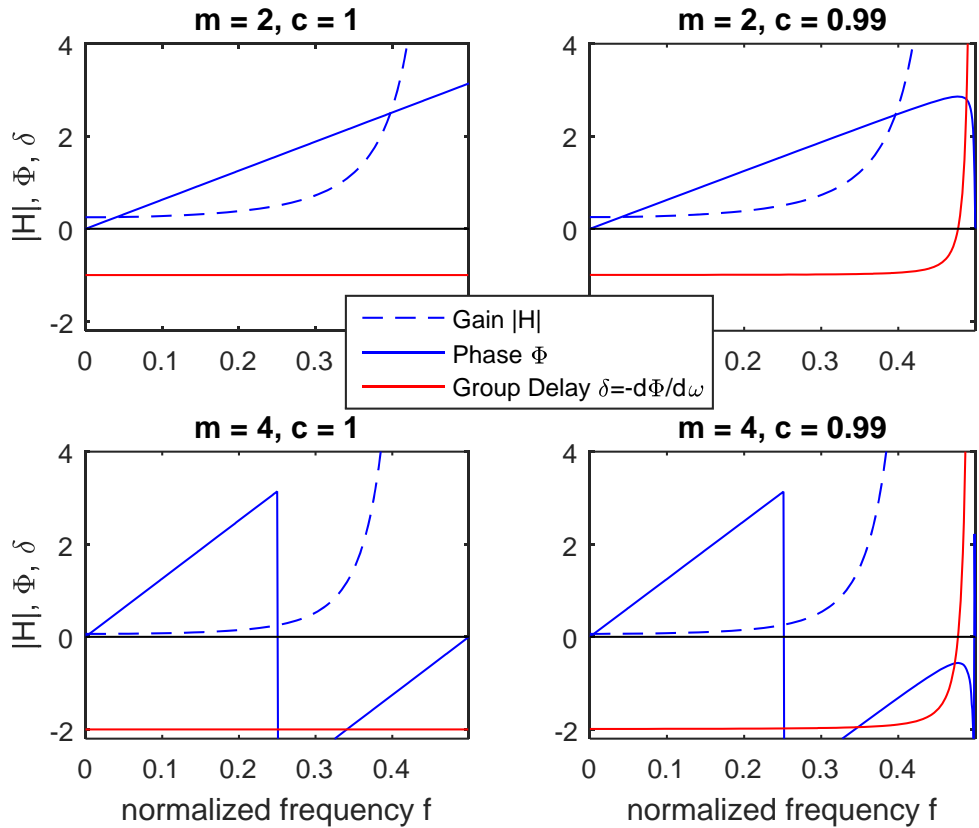

Figure 6: **Theory II – Cascading.**
Gain and phase of frequency response function (6) as well as corresponding group delay (7) for the cascaded system with τ = 1 and cascading level m = 2 (top row) and m = 4 (bottom row), for feedback gain c = 1 (left column) and c = 0.99 (right column). Note that although the NGD is constant over a wide frequency range, the gain increases for high frequencies, restricting applicability of the DINGD predictor to signals containing only lower frequencies. This is demonstrated in the example Figure 7.

Figure 7: **Chirp signal II – Cascading four predictors.**

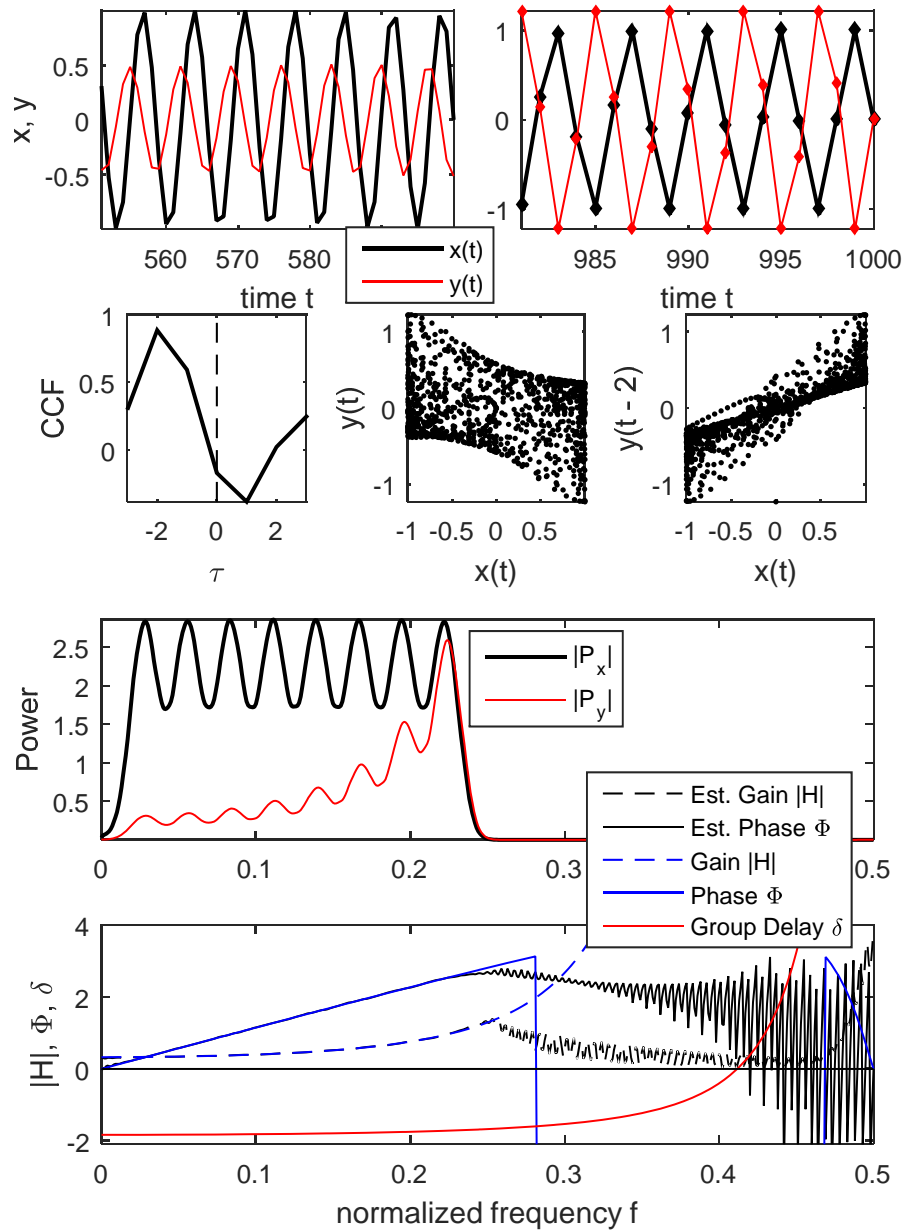

First row: The left plot shows 50 data points of a chirp signal x(t) and its prediction signal y(t). The right plot shows the last 20 data points. The DINGD predictor predicts the signal with the same group delay of -2, for both low and high frequencies. This is an example for a non-stationary signal that still can be predicted in real time.

Second row, left: The cross-correlation function has a peak value of CCF(-2) = 0.89. This shows that y(t-2) ≈ x(t), or, equivalently, y(t) ≈ x(t+2). Therefore, y(t) is a predictor of x(t). Center and right: The scatterplots confirm that y(t) is more correlated with x(t+2) than with x(t).

Third row: Power spectral density function estimates for x(t) and y(t). It is evident that frequency components of the signal that are closer to the resonance of the frequency response function, which equals the Nyquist frequency 0.5, are amplified more relative to lower frequency components. This explains the relatively smaller amplitude of the predictor for low frequencies.

Fourth row: Gain and phase of the frequency response function as well as corresponding group delay. Shown are analytic values from Eqs. (6) and (7) as well as gain and phase values estimated from the data, as shown in the legend. The large estimation errors of gain and phase for frequencies > 0.25 are due to the lack of input signal power for those frequencies.

Signal and prediction parameters: The signal x(t) consists of 1000 samples of a linear chirp signal with cutoff frequency of 0.25. Note that the chirp frequency sweeps a larger range than in Figure 3, made possible by the move of the resonance to higher frequencies. The parameters of Eq. (5) are $m = 4$, $b = 1.40$, $c = 0.85$, and $\tau = 1$.

## Shaping the frequency response by multiple delays

Cascading two DINGD systems corresponds to a total delay that is larger than the delays of the subsystems. Similarly, one could try to shape the frequency response function by using more than one delay in a single, non-cascaded, system. There are many possibilities, and here only three special cases with two delays and useful prediction properties are presented: A highpass, a lowpass, and a bandpass system with NGD. Rather than providing a detailed analysis, which would not add much to the already obtained insights, these systems are just stated, their frequency response functions are shown, and they are tested on numerical examples. The DINGD system with two delays is given by

$$y(t) = bx(t) - c_1 y(t - \tau_1) - c_2 y(t - \tau_2). \tag{8}$$

The coefficients are assumed to be non-zero and the delays are integers ≥ 1, as before for the basic DING predictor. The two-delay frequency response function is

$$H(\omega) = \frac{b}{1 + c_1 e^{-i\omega\tau_1} + c_2 e^{-i\omega\tau_2}}.$$

With

$$\beta(\omega) = 1 + c_1^2 + c_2^2 + 2c_1 c_2 \cos(\omega(\tau_1 - \tau_2)) + 2c_1 \cos(\omega\tau_1) + 2c_2 \cos(\omega\tau_2)$$

it follows

$$|H(\omega)| = \frac{|b|}{\sqrt{\beta(\omega)}}$$

and

$$\delta(\omega) = -\frac{d\Phi(\omega)}{d\omega} = -\frac{c_1^2 \tau_1 + c_2^2 \tau_2 + c_1 \tau_1 \cos(\omega\tau_1) + c_2 \tau_2 \cos(\omega\tau_2) + c_1 c_2 (\tau_1 + \tau_2) \cos(\omega(\tau_1 - \tau_2))}{\beta(\omega)}.$$

• Lowpass system with $c_1 = 2$, $c_2 = 1$, $\tau_1 = 1$, $\tau_2 = 2$. The phase, gain, and group delay are shown in Figure 8, first row. This system has a group delay of -1. The numerical simulation data in Figure 9 has a low-frequency sinusoidal added to a lowpass filtered noise signal in order to demonstrate that low frequency components do not detrimentally affect prediction of the high-pass components. The CCF, validating the group delay of -1, looks quite remarkable.

• Highpass system with $c_1 = -2$ and otherwise same parameters as before. This system has a group delay of -1. The phase, gain, and group delay are shown in Figure 8, second row. A numerical example is provided in Figure 10.

• Bandpass system with $c_1 = -2$, $c_2 = 1$, $\tau_1 = 2$, $\tau_2 = 4$. The phase, gain, and group delay are shown in Figure 8, third row. This system has a group delay of -2. To use this system instead of the low- or highpass systems can have additional advantages in additon to the higher NGD; the behavior of the gain allows for somewhat higher/lower frequency components of the signal as in the lowpass/highpass systems. Also, compared with the single-delay system in Figure 4, this two-delay system allows for a larger signal bandwidth. This is shown in the numerical example of Figure 11. Of all examples, the signal here appears to be the most complex one. That this signal can be predicted in real time two time points ahead by the simple mechanism of delay-induced negative group delay, Eq. (8), is not trivial.

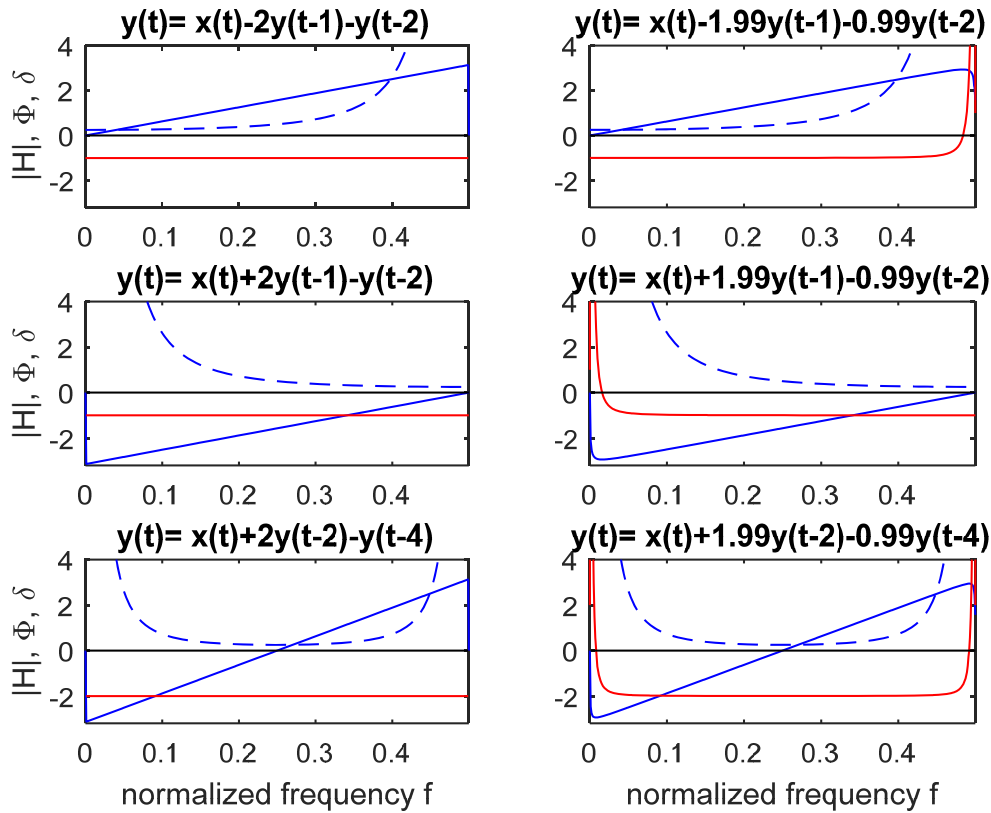

Figure 8: **Theory III - Multiple delays**.
First row: Lowpass DINGD system with group delay of -1. Shown are gain and phase of frequency response function as well as corresponding group delay for the lowpass system with parameters stated in the text and also provided in the figure titles ($b = 1$). Again, the left column has theoretical parameters that cause a constant NGD for all frequencies, and the right column has more realistic parameters. (The graphs in the left column are identical to the graphs of the cascaded system in Figure 6 top row, left column)).
Second row: Highpass DINGD system with group delay of -1.
Third row: Bandpass DINGD system with group delay of -2. The two delays are now twice as large as before, i.e., $\tau_1 = 2$ and $\tau_2 = 4$. For legends, please refer to Figure 6.

## Figure 9: Lowpass DINGD system.

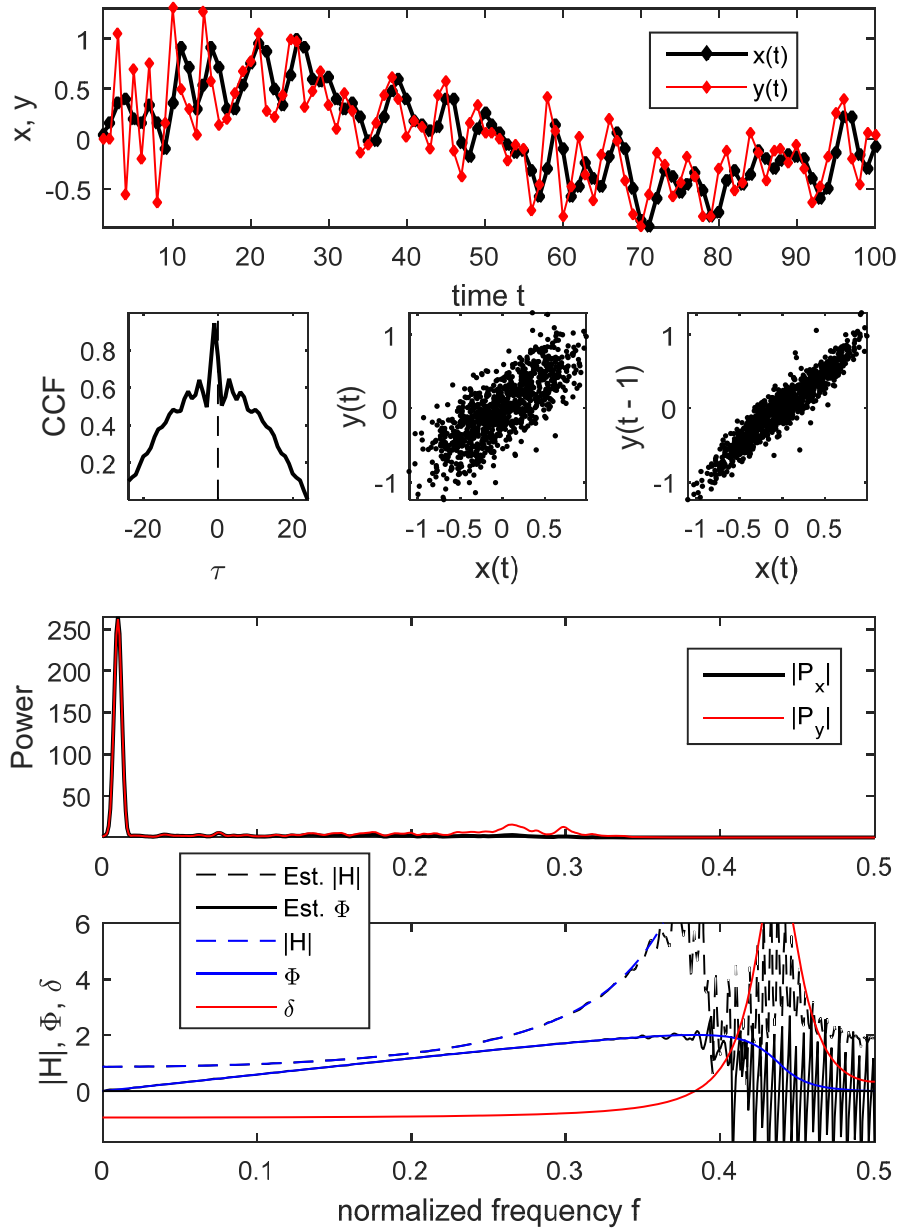

First row: The first 100 data points of a lowpass filtered noise signal x(t) with an added sinusoidal signal with frequency 0.01 and its prediction signal y(t). Due to the very low frequency component of the sinusoidal, relatively long transients of the prediction signal are visible for small t.

Second row, left: The cross-correlation function has a peak value of CCF(-1) = 0.94. This shows that y(t-1) ≈ x(t), or, equivalently, y(t) ≈ x(t+1). Therefore, y(t) is a predictor of x(t). Center and right: The scatterplots confirm that y(t) is more correlated with x(t+1) than with x(t).

Third row: Power spectral density function estimates for x(t) and y(t).

Fourth row: Gain and phase of the frequency response function as well as corresponding group delay. Shown are analytic values as provided in the text as well as gain and phase values estimated from the data, as shown in the legend. The large estimation errors of gain and phase for frequencies > 0.35 are due to the lack of input signal power for those frequencies.

Signal and prediction parameters: The signal x(t) consists of 1000 samples of white noise, low-pass filtered with a Butterworth filter of seventh order with a cutoff frequency of 0.30 with an added sine signal of amplitude 0.5 and frequency 0.01. The parameters of Eq. (8) are b = 3.00, $c_1$ = 1.65, $c_2$ = 0.80, $\tau_1$ = 1, $\tau_2$ = 2.

Figure 10: **Highpass DINGD system.**

First row: A highpass filtered noise signal x(t) and its prediction signal y(t). Out of 1000 simulated time points, 50 are shown. Similar to Figure 4, it is evident that patterns of data points that are not formed by the envelope of an oscillatory signal are predicted well.

Second row, left: The cross-correlation function has a peak value of CCF(-1) = 0.89. This shows that y(t-1) ≈ x(t), or, equivalently, y(t) ≈ x(t+1). Therefore, y(t) is a predictor of x(t). Center and right: The scatterplots confirm that y(t) is more correlated with x(t+1) than with x(t).

Third row: Power spectral density function estimates for x(t) and y(t).

Fourth row: Gain and phase of the frequency response function as well as corresponding group delay. Shown are analytic values as provided in the text as well as gain and phase values estimated from the data, as shown in the legend. The large estimation errors of gain and phase for frequencies < 0.15 are due to the lack of input signal power for those frequencies.

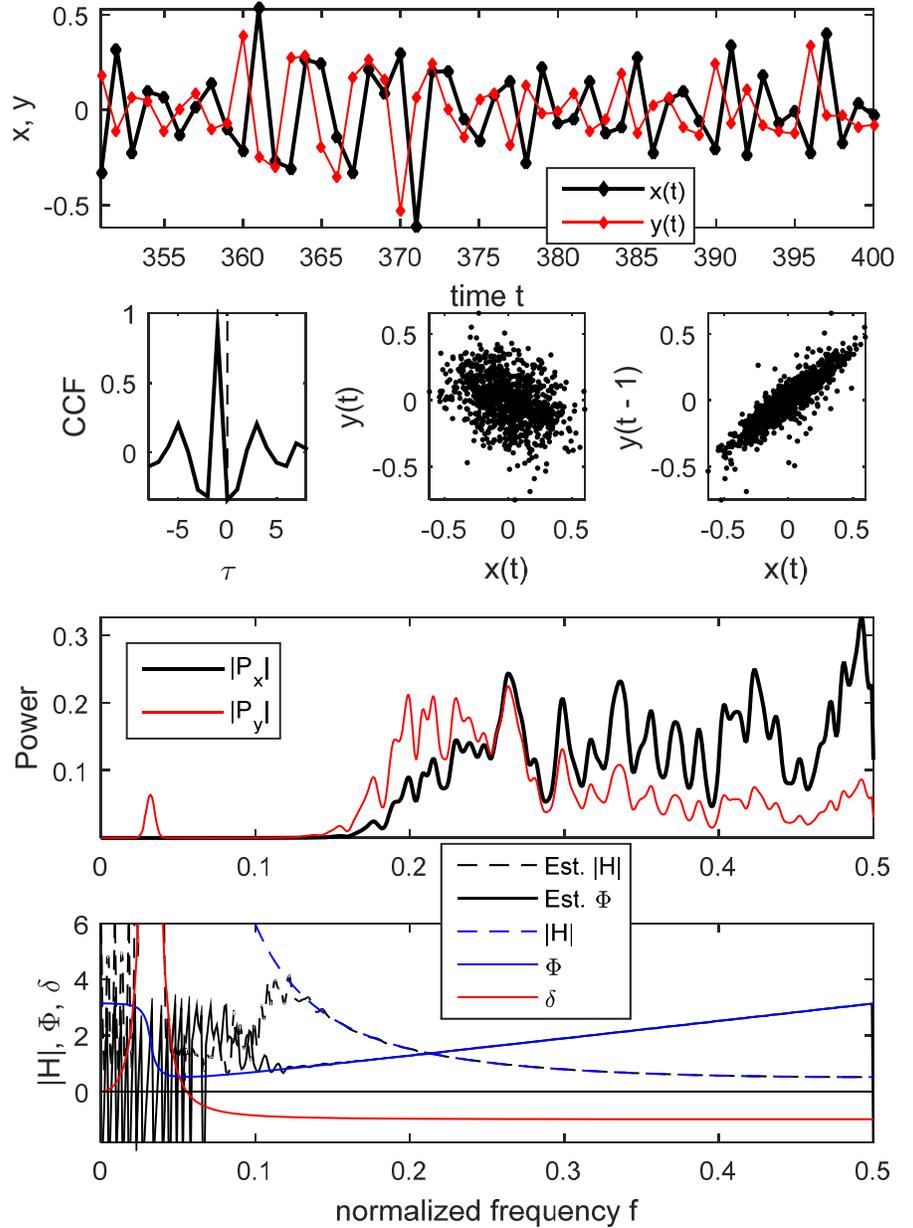

Signal and prediction parameters: The signal x(t) consists of 1000 samples of white noise, low-pass filtered with a Butterworth filter of seventh order with a cutoff frequency of 0.20. The parameters of Eq. (8) are b = -2.00, $c_1$ = -1.92, $c_2$ = 0.96, $\tau_1$ = 1, $\tau_2$ = 2.

**Figure 11: Band-limited noise III – Bandpass DINGD system with group delay of -2.**

First row: A bandpass filtered noise signal x(t) and its prediction signal y(t). Out of 1000 simulated time points, 50 are shown.

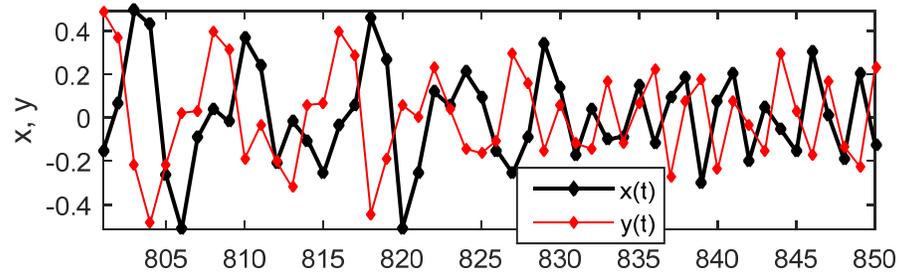

Second row, left: The cross-correlation function has a peak value of CCF(-2) = 0.90. This shows that y(t-2) ≈ x(t), or, equivalently, y(t) ≈ x(t+2). Therefore, y(t) is a predictor of x(t). Center and right: The scatterplots confirm that y(t) is more correlated with x(t+2) than with x(t).

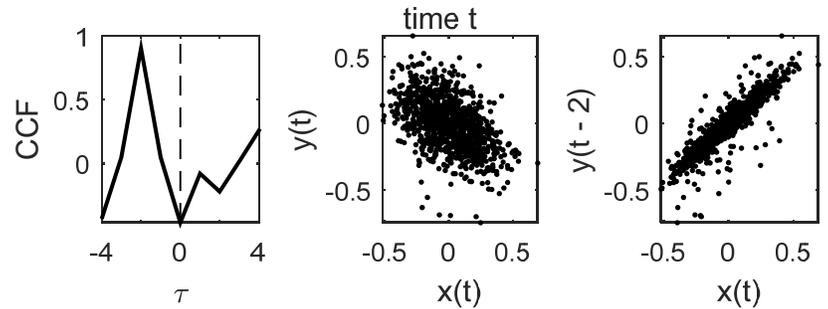

Third row: Power spectral density function estimates for x(t) and y(t).

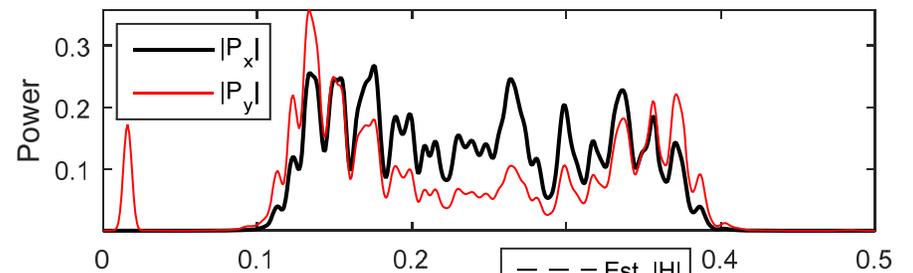

Fourth row: Gain and phase of the frequency response function as well as corresponding group delay. Shown are analytic values as provided in the text as well as gain and phase values estimated from the data, as shown in the legend. The large estimation errors of gain and phase for frequencies between 0.1 and 0.4 are due to the lack of input signal power for those frequencies. Remarkably, the group delay is constant -2 for all frequencies.

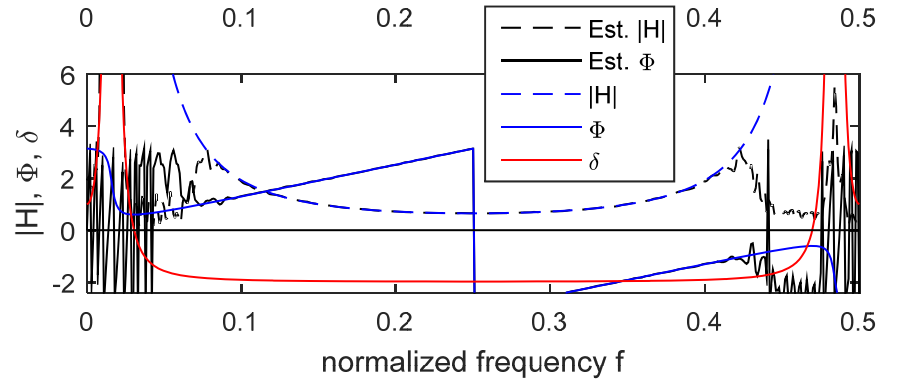

Signal and prediction parameters: The signal x(t) consists of 1000 samples of white noise, bandpass filtered with a Butterworth filter of seventh order with a cutoff frequencies of 0.12 and 0.38. The parameters of Eq. (8) are b = -2.50, $c_1$ = -1.90, $c_2$ = 0.94, $\tau_1$ = 2, $\tau_2$ = 4.

The MATLAB code for generating this figure is available from the author on request.

# Conclusions and discussion

A very simple discrete-time predictor based on delayed feedback-induced NGD has been described. This delay-induced NGD (or DINGD) predictor predicts future signal values by using present input and past output, or already predicted, signal values. It thus differs from most other predictors or forecasting models, which do not take into account past predicted but only past *input* signal values [24, 25]. (An exception would be systems that use anticipatory synchronization for prediction [26]). In other words, most conventional time series predictors can be written in the form

$$\text{Conventional predictor: predicted } x(t+1) = f(x(t), x(t-1), x(t-2), \ldots),$$

in which f(.) is a specific model of the time series to be predicted. These conventional predictors depend on prior observations but not prior predictions, and usually contain coefficients resulting from a fit to a fixed learning data set or which are continuously being updated [27]. In contrast, the DINGD predictor cannot be written in this explicit form, as it depends also on already predicted values. For example, all DINGD predictors with prediction horizon of 1 considered above can be written as

$$\text{DINGD predictor: predicted } x(t+1) = y(t) = g(x(t), y(t-1), y(t-2), \ldots),$$

where g(.) is a function depending on the spectral content of the time series to be predicted. This expression is fundamentally different to the analogous expression for conventional prediction above. For example, Eq. (1) with τ = 2 and c = 1, when the group delay is -1 and the prediction horizon equals 1, could be written as

$$\text{predicted } x(t+1) = y(t) = bx(t) - y(t-2).$$

It is apparently not possible in this case to express the predicted x(t+1) in closed form as a function of a finite number of past values of x(t) only. Rather, the DINGD predictor resembles a dynamic form of prediction, also called anticipatory relaxation dynamics [14], than conventional prediction. It is this dynamic origin of prediction that makes the DINGD predictor, and possible other NGD-based prediction, so counterintuitive. However, avoiding past input values might be advantageous for prediction by natural and artificial neuronal networks as only already predicted, internalized states need to be laid down in memory.

In addition, the DINGD predictor does not require a learning data set but is a true real-time predictor once the coefficients have been fixed and the dynamics have settled into a steady state. As the non-stationary chirp signal examples have indicated, it is worth looking into the real-time advantages of this prediction scheme compared to adaptive prediction algorithms as well. Furthermore, it has been demonstrated here that the DINGD predictor is able to predict complex non-smooth signals, too, adding insights beyond the contemporary understanding of NGD [7, 10, 28].

It would be interesting if DINGD predictors could be applied or physically implemented in some way. Of particular interest could be the fact that only past predicted signals are used for prediction. It means for example for a predictive agent such as a robot [29] or the brain [30, 31] that only internal state variables are needed for prediction, without having to store sensory inputs in memory. Further, the main ingredient for the DINGD predictor are time delays, which are abundant in the central nervous system including the cerebellum [32, 33].

## Acknowledgment

I would like to thank Blaise Ravelo for discussions.